# Laudatores Temporis Acti, or Why Cosmology is Alive and Well – A Reply to Disney

Milan M. Ćirković
Astronomical Observatory, Volgina 7, 11160 Belgrade, YUGOSLAVIA
E-mail: arioch@eunet.yu

*All science is cosmology, I believe.*

Sir Karl Popper

*To search thoroughly for the truth involves a searching of souls as well as of spectra.*

Georges Lemaître

**Abstract.** A recent criticism of cosmological methodology and achievements by Disney (2000) is assessed. Some historical and epistemological fallacies in the said article have been highlighted. It is shown that—both empirically and epistemologically—modern cosmology lies on sounder foundations than it is portrayed. A brief historical account demonstrates that this form of unsatisfaction with cosmology has had a long tradition, and rather meagre results in the course of the XX century.

## 1. Introduction

In a recent intriguing and provocative paper, Disney (2000) has suggested that our cosmological knowledge and contingent claims are "overblown", as well as "naïve in the extreme", "doing a disservice", etc. There are other recent instances of similar criticism being expressed at scientific conferences, publications and informal debates. It is the purpose of this essay to show that such criticism—although certainly not unwelcome—is not completely fair, it does leave several important facts and issues out of the picture, and that it is not particularly new or original. These considerations dictate the division of the presented material: after we consider criticisms of the empirical basis of modern cosmology (Section 2), we shall tackle the deeper issues of the epistemological foundations of the contemporary cosmological work (Section 3). Before concluding this reply, we shall demonstrate that Disney's criticisms are mainly rehashed versions of the arguments appearing already several times in the history of the XX-century science.

The immediate motivation for the Disney's critique is the claim in the recent paper of Hu, Sugiyama and Silk (1997) to the effect that with the information to be supplied by the new generation of the cosmic microwave background (henceforth CMB) satellite observatories we shall be able to put the standard Big Bang cosmological model on the same epistemological level as the Standard Model of particle physics. Creating a controversy out of such optimistic view, Disney pronounces:

> We believe the most charitable thing that can be said of such statements is that they are naive in the extreme and betray a complete lack of understanding of history, of the huge difference between an observational and an experimental science, and of the peculiar limitations of cosmology as a scientific discipline. By building up expectations that cannot be realised, such statements do a disservice not only to astronomy and to particle physics but they could ultimately do harm to the wider respect in which the whole scientific approach is held. As such, they must not go unchallenged.

In the further course of this note, we shall show that it is exactly such broad criticism of cosmology articulated by Disney which is rather naive and which demonstrates a lack of understanding of history of science. Furthermore, one can easily show that the very essence of such criticism is almost verbatim repetition of various occurences in the history of cosmology. Inverting the famous remark of George Santayana, one may say that those who relive the past thus betray their lack of historical knowledge.

First of all, the broad optimistic expectations of proponents of CMB experiments are not alone in the entire spectrum of physical sciences in comparing the relativistic cosmology with our contemporary models in particle physics. Let us mention here just one famous example. According to the amusing, but remarkable, classification of Penrose (1989, p. 152), our theories in natural sciences may be divided into three broad categories of SUPERB, USEFUL and TENTATIVE. While the Big Bang cosmology is put by Penrose into category of USEFUL theories, so is the Standard Model of particle physics (that is, its two main constituents, the quark model and the electroweak theory of Glashow, Salam and Weinberg)! Therefore, the statement of Hu et al. (1997) can hardly be regarded as something outlandish and senseless, as Disney would have us believe.

There is, however, a very significant and important point that Disney's and similar criticisms are signposts of. It may be explicated as insufficient attention devoted to epistemological and other philosophical aspects of modern cosmology, as compared, for instance to biological sciences or quantum mechanics. In particular, when the rate of growth of knowledge is taken into account, it is clear that philosophical foundations of modern cosmology are still very inadequate, its textbook framework often outdated or directly wrong, and misconceptions about it should not be surprising. This point (and further motivation for engaging in the present discussion) is described in wise words of Kragh (1997) that

> there is a tendency to streamline history and ignore the many false trails and blind alleys that may seem so irrelevant to the road that led to modern knowledge. It goes without saying that such streamlining is bad history and that its main function is to celebrate modern science rather than obtain an understanding of how science has really developed. The road to modern cosmology abounded with what can now be seen were false trails and blind alleys, but at the time were considered to be significant contributions.

In reminding us of this lies the true value and importance of Disney's (2000) study.

## 2. Empirical basis of modern cosmology

Disney lists 13 observations constituting, in his opinion, the entire empirical basis of cosmology (Table 1 of Disney 2000). There are several problems with this list. The most obvious one is the lack of several important observational data clearly relevant to cosmology. Lacking are, for instance, observations on the properties of galaxy clusters, especially those at high redshift, as well as appropriate Sunyaev-Zel'dovich effects (e.g. Sunyaev and Zeldovich 1980).[1] Those have already showed their potential to discriminate between various cosmological models, in particular to reject the class of steady-state models with continuous recycling of high-entropy matter (Peter Phillips, personal communication). Beside abundances of deuterium and helium cited, the abundance of lithium is also relevant to cosmology (Deliyannis and Demarque 1991; Deliyannis et al. 1994). However, the most glaring omission is the very existence of high-redshift objects, especially after the small revolution brough about by HDF and subsequent deep-field observations (e.g. Lanzetta, Yahil and Fernandez-Soto 1996), which places significant constraints on the various cosmological models of structure formation. Among many important contributions in this field, it should not be forgotten that Disney himself has contributed much to our understanding of the high-redshift universe (e.g. Phillipps, Davies and Disney 1990). In addition, there are observations and results nominally belonging to other scientific disciplines which are obviously relevant to the modern cosmology. Notably, the existence of large-scale entropy gradient ("the arrow of time") is one such part of empirical knowledge employed in cosmology[2]; other examples include flavors and masses of elementary particles, or observational indications (from studies of binary pulsars) for the existence of gravitational waves. If we should take some lessons from the developments in various physical sciences in last several decades, we should conclude that there can be no "appropriation" of empirical data: it is at least suspicious to say that—in the tight network of the present day research—piece of data A belongs to exclusively cosmology, piece B to (say) galactic dynamics, and piece C to particle physics.

At several points, the Disney's discourse displays such implicit clear-cut distinction between cosmology and astrophysics. This in itself represents a claim very difficult to support. Where exactly is the boundary located? Does the formation of the large scale structure belong to the cosmological or "just" astrophysical domain? Formation of individual galaxies? Formation of the Milky Way? Numerical simulations which are criticized (and largely correctly so) by Disney do in fact deal with the hierarchical structure formation, the process which continues to this day (and thus, in principle, **is** observable)! If we discard the artificial boundary between cosmological and astrophysical domains, a whole new array of observational data becomes accessible and practically useful to cosmology. A nice example along this line of thought is the Gunn-Peterson test (in its various versions, both H I and He II), and the entire field of investigation of the early universe stemming from it (Gunn and Peterson 1965; Giallongo et al. 1994; Fang and Crotts 1995; Jakobsen 1997). By sharply delineating these fields, Disney used the opportunity to discard such (sometimes magnificently precise) observations and reduce the relevant list of empirical data. Obviously, this procedure is biased, to say at least.

Some items on the Disney's list are marked as "serendipitous", and in several other instance in the article is such serendipitous nature of many cosmological discoveries emphasized as if this somehow constitutes an argument against cosmology. We fail to see how this factographic background can be construed as derogatory. The history of **all** scientific disciplines is full of serendipitous discoveries, from the radioactivity to QSOs. If one wishes to be biased in this manner, one may also construe all the results following from such random occurences as comets and supernovae as serendipitous and somehow of lesser significance, since no amount of systematic effort (or "good laboratory empiricism" or verificationism) will bring about, for instance, SN 1987A. Therefore, such labelling is truly irrelevant to the issue whether cosmology is a proper science or not, or whether its database is big or small.

---

[1] The latter may be said to be included in "measured fluctuations in CBR". However, the question mark added suggests that the author intends to cite *intrinsic* fluctuations.
[2] That this is a cosmological datum was clear to Boltzmann and Zermelo at the end of XIX century. See, for instance, a vivid account of their polemic in Steckline (1983). Modern accounts of this all-important issue can be found in Zeh (1992) and Price (1996).

The fact that observations are different from experiments is rather well-known and hardly deserves much dwelling upon. However, on the basis of this distinction one can hardly discard observational disciplines as unscientific. It seems that Disney would not have felt so uncomfortable if somebody has published opinion similar to the one by Hu et al. (1997) that after we launch the next generation of interplanetary spacecraft "we shall know more about satellites of Jupiter than about quarks", in spite of the obvious fact that our entire corpus of knowledge on the Jovian system is of observational nature. The difference between this knowledge and the contemporary cosmological knowledge, therefore, is mostly psychological in nature. On the other hand, it is highly overblown to claim that the Standard Model of particle physics is based upon "thousands of experiments". Even if one can count thousands of individual experiments (which is unlikely to be possible without massive historical research, not supported in Disney's article), it only means that all historical observations of, say, CMB should be counted (and there are certainly "dozens" if not "hundreds" of them). In particular, since we should count the arrow of time as a cosmological datum, the way of comparison of disciplines implied by Disney is reduced to absurd, since one may say that there are literally billions of pieces of empirical data relevant to cosmology. The real question is, on the other hand, how many different kinds of experiments or observations may independently be checked against existing theoretical models. And when the situation is considered in this light, one may see that the situation in the contemporary particle physics is not much better than in the contemporary cosmology (or in any other frontier discipline of science as well), and that it closely corresponds to the Penrose's classification, rather than the view presented by Disney.

Finally, there is a point of practical significance in the years to come. It does not seem fair (if not outrightly hypocritical), to complain of the small cosmological database, simultaneously shooting poisoned arrows at that part of the cosmological community attempting to motivate the Next Generation Space Telescope. It is strange to expect that we may be ever "confident of converging on 'the truth'" and still profess knowledge of the history of science. There has been no such epistemological "convergence" in history; it was always just hard searching.

## 3. Epistemological fallacies in criticism of cosmology

What does motivate the strange (but, as we shall see in the next section, rather old-fashioned) pronouncement that "this fascinating subject... is not a proper science at all"? This claim is repeated at least three times in the course of rather short Disney's article. In light of the seriousness of this charge and its implication to both academic life and teaching science, its all aspects should be investigated in as much detail as possible.

"There is only one Universe!", exclaims Disney. Although this can also be regarded as an article of faith (and there are some reasons in modern quantum cosmology to believe otherwise), let us accept it for a moment and take a look at its epistemological consequences. In several instances, Disney suggests that the status of cosmology as a science is somehow jeopardized by the fact that there is only a single universe. *Per analogiam*, it is highly uncertain whether evolutionary biology is a science (there has been only one, unrepeatable evolution of terrestrial biosphere), whether history is a science (unrepeatable sequence of historical events), whether geology is a science (only one Earth!), whether archaeology is a science (only a single array of material traces of ancient cultures), etc. However, if we are more open-minded, especially in the spirit of Sir Karl Popper's quotation from the beginning of the present paper, we shall realize that it is quite a common situation in science. Even more suspicious is the following locution:

> At a stroke this removes from our armoury all the statistical tools that have proved indispensable for understanding most of astronomy.

A man in the street, not knowing anything about cosmology, could think (after reading Disney's verdict) that there is no statistics in cosmology. This is simply wrong. There are dozens of instances (from N-body simulations of structure formation, to the statistics of clusters and voids, to the CMB peaks) of usage of both statistical apparatus and the statistical mode of thinking in modern cosmology. Disney's remark is simply off the mark. The fact that one can not empirically investigate *ensemble of universes* is in itself an empirical fact and presents a part of our knowledge which we take into account in all cosmological considerations. It is not something extraneous which cosmologists simply ignore.

Further remarks pertain to extrapolations required by the cosmological discourse. This is a celebrated issue, which has worried cosmologists from the beginning of this century. It has, on the historical side, represented one of the major motivations for building the steady-state cosmological model, especially in the version of Bondi and Gold (1948). Many physical and philosophical discussions have taken place since that time, and many important issues have been cleared. However, it still worries people, as testified by Disney, who writes:

> Cosmology requires us to extrapolate what physics we know over huge ranges in space and time, where such extrapolations have rarely, if ever, worked in physics before. Take gravitation for instance. When we extrapolate the Inverse Square Law... from the solar system where it was established, out to galaxies and clusters of galaxies, it simply never works. We cover up this scandal by professing to believe in "Dark Matter" – for which as much independent evidence exists as for the Emperor's New Clothes.

It needs to be noticed that statements like this hardly testify on the pretended "understanding of history", since situation similar to the one described (in rather strong terms) is often a common working situation in science. Let us take an example exactly from particle physics, which Disney implies to be a real, mature and serious scientific discipline. At any time between cca. 1930. and 1960. a Disney-like sceptic could write a damaging paper against the nascent particle physics containing the following locution:

> Particle physics requires us to extrapolate what physics we know over minuscule ranges in space and time, where such extrapolations have rarely, if ever, worked in physics before. Take conservation laws for instance. When we extrapolate the laws of energy and momentum conservation...

from the mechanics and thermodynamics where they were established, down to the world of radioactivity and in particular the ubiquitous beta-decay, they simply never work. We cover up this scandal by professing to believe in "neutrinos" – for which as much independent evidence exists as for the Emperor's New Clothes.

Although a further elaboration of the analogy is hardly necessary, one should repeat here that whoever preaches humility should first realize its complete implications. It is with utmost pretentiousness that one concludes that there should be no stable dark matter particles only because we haven't found positive evidence for one in a very, very tiny range of time, space and detectability characterizing our experimental devices. In addition, as Bondi (1962) correctly put it, "We know from many directions that it is a very dangerous principle to suppose that anything we have in physics holds with arbitrary accuracy."
We find ourselves in similar position regarding complaints such as:

> Objects at cosmologically interesting distance are exceedingly faint, small and heavily affected by factors such as redshift-dimming and K-corrections, so it will obviously be very difficult, if not impossible, to extract clear information about geometry, or evolution, or astrophysics – all of which are tangled up together.

The factographic background of this statement is impeccable. However, **exclusivity** is the real issue here: the same may be said for almost any other frontier of science. How about

> Objects and events at scales characteristic for grand unified theories are too tiny, too fuzzy, too rare and too difficult to interpret, being affected by uncertainties in cross-sections, rare decays, ... so it obviously will be very difficult, if not impossible, to extract clear information about physics at those scales.

Now, it is interesting that the very terminology and the very reasons Disney lists (redshift-dimming, K-corrections, etc.) are **inseparable parts of the cosmological repertoire**! In other words, only by improving our **cosmological** knowledge we may improve the database, which is, of course, quite normal self-correcting procedure in all scientific disciplines. there is no peculiarity of cosmological epistemology in this regard.
Interestingly enough, Disney sees "very little evidence" to support, among others, the assumption that the portion of the universe susceptible to observations is representative of the cosmos as a whole. But this is the central thesis of the entire Copernican revolution, and whoever doubts it may as well affirm Aristotelian belief in immutable heavens (cf. the discussion of Dingle's criticisms below)! This can hardly be called "luck" as Disney tends to do. Together with other assumptions which Disney attaches to cosmology (and which are highly controversial from the beginning to the end), this is intended to show that cosmology is very limited discipline. However, a little bit of careful thinking would indicate that similar set of assumptions can as well be applied to particle physics or genetics or indeed any other modern, rapidly advancing field. Moreover, some modern concepts, like the anthropic principle and various inflationary scenarios, do come very close to relaxing this assumption. There are almost self-evident reasons, embodied in the Weak Anthropic Principle, as defined by Carter (1974), why we do expect to see the universe severely limited by restriction imposed by our very existence; it is completely different and very intriguing question (which is certainly beyond the scope of the present article) what physical **agencies** of such fine-tuning may be, although here also intriguing hypotheses are not lacking (e.g. Smolin 1992).

The complexity issue is also necessary to tackle here, since it occupies a prominent position among Disney's criticisms. "The optimistic cosmologist can always counter argue (I don't know how) that the Universe in the large is a great deal simpler than its constituent parts." Historically speaking, it is in fact very easy to argue in that direction. It is clear that the structure of atom, with a nucleus regarded as a point of given mass and charge, is sufficient to explain all phenomena in atomic physics, chemistry, thermodynamics, etc. On the other hand, this simplification is not warranted on a deeper level, dictated by quantum mechanics. A simple historical study of Dalton's chemistry vs. Mendeleev's chemistry vs. London's chemistry will show that. In non-linear systems, such as our universe certainly is, the phenomenon that "the whole is less than its parts" is quite a common feature. Even in simplest mathematics, there are countless such cases. Stock examples include the uniform probability distribution: a specification of a real point and corresponding value of probability requires an infinite amount of information. At least one prominent author (Tegmark 1996) has argued on that basis that the entire universe (or, more to the point, **multiverse**) is built in the same way: with negligible total algorithmic informational content, and all apparent information contained in, say, our astronomical catalogues, being just a illusion due to the decoherence.

Many of these issues have been answered, among others, by Sir Hermann Bondi in his illuminating Halley Lecture (Bondi 1962), as well as in a later booklet (Bondi 1967). In a lively and colorful discourse, Bondi stated, in connection with the relationship of Michelson-Morley experiment, aether and the Special Theory of Relativity:

> Naturally, this is an example where you see, with a great deal of hindsight, how much simpler it would have been to have thought the matter through first instead of performing the experiment, but this naturally is a totally unhistorical approach. We can think it through so simply now and come to such a clear cut result, because our habits of thought have been formed by the insight due to this experiment. This has been, and is, a very valid criticism of every approach in which we try to account for complicated features of our environment found, often with great difficulty, by experimental methods, in terms of some quite simple and elementary fact which we then think through to its bitter conclusion, and scientists, rightly, are always very suspicious of hindsight... It is a good thing to consider as far as we can, which is usually not very far, the various possibilities, so that at least we can try to formulate crucial questions.

This applies as well to Hu et al.'s and Disney's comparison of cosmology and particle physics. For other issues of epistemological relevance, we direct the reader to Bondi (1967) booklet, since it would be too straightforward to list the relevant arguments here.

Finally, we have those issues in Disney's critique which are basically sociological in nature and thus are irrelevant for the criticism of cosmology *per se*. This applies, for instance, to the complaints that cosmology is "unhealthily self-referencing", repetitive and devoid of genuine scientific debate. It is extremely doubtful whether there is any discipline in modern sciences to which such problematic attributes can not be assigned after some selective research. What about much discussed particle physics, whose obvious successes justify its choice as a standard for physical sciences? Is the "dogma" of Cold Dark Matter any worse than the "dogma" of Grand Unified Theory or even something called The Theory of Everything (which is firmly supposed to exist, but there is not a trace of agreement what it should look like, how many free parameters contain, on what energies, etc.)? And what about all myriads of predicted particles, all those axions, photinos, gravitinos, arions, sneutrinos, various Higgs' particles, etc. about which countless papers, books and dissertations have been written without any empirical support whatsoever—but very thick ("self", and of necessity) referencing? The true problem does not lie in cosmology; it is, if anywhere, in the heart of modern understanding of science, together with citation impacts, tenure-hunting and other parafernalia which form a legitimate subject for a critical sociology of science—but we certainly can not enter into that area here.

## 4. Laudatores Temporis Acti: good old cosmology bashing

In his famous *Book of Imaginary Beings*, Jorge Luis Borges describes a sect or group of people devoted to celebrating and re-living the past, appropriately called Laudatores Temporis Acti (Borges 1970). Borges himself judges their belief as senseless and unproductive. The same may be said for a long-standing tradition of criticism of scientific nature and epistemological grounding of cosmology. If history bears any relevance, than such attacks can not result in any advance of our knowledge.

Since Disney listed among cosmology's failures pre-physical, that is, religious cosmologies of the ancient times, it would only be just to reply with the historical fact that essentially cosmological issues (of course, non-orthodox) were among the most repressed scientific achievements, in particular in Europe during the Middle Ages and the Rennaisance. Wasn't Bruno burnt at stake exactly for proposing a dissenting cosmological view (of the plurality of worlds)? The repression of ecclestiacal authorities against the cosmological revolution of the XVI and early XVII century is a well-known story. However, since we maintain that the true subject of inquiry should be only *physical cosmology*, we should consider only criticisms arising within that framework (that is, since 1916. and the advance of General relativity).

Thus Bishop Barnes, one of the most distinguished natural philosophers in the first several decades of XX century wrote that the Lemaitre's "primeval atom" hypothesis (that is, the first Big Bang cosmology as understood today) is "a brilliantly clever *jeu d'esprit* rather than a sober reconstruction of the beginning of the world" (Barnes 1933). Obviously, to this statement, written in 1933, Disney and other modern detractors of cosmology may fully subscribe.

Historically the most interesting attacks on cosmology came during the two debates, in late 1930-ies and in 1950-ies, initiated by Herbert Dingle. In the first wave of attack, Dingle (1937a,b) criticized cosmological theories of Eddington, Milne and Dirac, which enjoyed rather great popularity and publicity in these days (which is testified by replies ot Dingle's papers in the Letters section of *Nature*, among others by Whitrow, McCrea, Haldane and C. G. Darwin). The essence of his attack was that those cosmologists (although they have not been labeled as such at the time) perverted empirical basis of natural sciences, being too rationalistic, extravagant and even mysticism-prone. In a pejorative sense, he etiquetted his targets even in the very title of his article as "modern Aristotelians" and accused them of pseudo-religious "cosmolatry" (compare Disney's *dictum* that "the most unhealthy aspect of cosmology is its unspoken parallel with religion"!). Ironically enough, Dingle censures Eddington for believing that the cosmological constant $\Lambda$ is of fundamental importance in nature, the belief that has been largely vindicated in recent years (cf. Krauss and Turner 1999). Basic thrust of his attack was the same as Disney's: cosmological theories are speculative, empirically unsupported and highly dogmatic.

In the next round of his battle with cosmology, a decade and a half later, Dingle discovered new and more modern (as well as dangerous) targets: the steady-state cosmologists, mainly Bondi and Hoyle (Dingle 1953, 1956). Incidentally, they were targets of the criticisms of most of the mainstream cosmologists, but mainly for reasons entirely different from Dingle's (Kragh 1996). The general tone of Dingle's criticism in 1950-ies was the same as in 1930-ies: cosmology is not a true science, it can not be found on any general principle, only inductive results are permissible, empiricism is the correct scientific attitude, everything else is Aristotelianism and religion. It goes without saying that Dingle rejected the relativistic cosmology (i.e. all Big Bang models) for these same reasons, although, obviously, the steady-state camp was an easier target.

It is not necessary to emphasize that the Dingle's criticisms were, although certainly well-intentioned, unfair and far off the mark; they share both properties with the Disney's critique. Theories like Dirac's, Milne's and the steady-state were certainly not mystic nor extravagant; the former is applicable rather to their opponents, and the latter is not something which can be objectively determined, but rather the issue of habits and fashion. Narrow inductivism and a particularly strong and unproductive version of verificationism underlie all such attacks and transform them into a sort of *ignorabimus*-festival.[3] The plain fact that our knowledge is often unreliable does not provide, in itself, the ever-lasting topic for productive discussion. The best general medicine in these cases is the very history of science. It is enough to ask: could we obtain great, successful, uncontroversial scientific theories by such strait and inflexible approach? Are epistemological bases of, say, Darwin's theory of evolution, Maxwell's electrodynamics or Einstein's General relativity compatible with extreme inductivism and verificationism advocated by Dingle and Disney? The answer seems obviously negative. The accusation of "cosmolatry" is difficult to answer, being so subjective and emotionally-charged. In any case, it seems possible to find pathologies in worshipping of experiment and empirical knowledge as well.

Similar criticisms have been repeated almost verbatim in the Western world by Munitz, Bunge, Zwicky, Scriven and other authors (interestingly enough, mainly philosophers of science, or observational astronomers with philosophical ambitions, like Zwicky). Simultaneously, the basically same issues have served as ready ammunition for Stalinist attacks on cosmology (probably the second most repressed scientific discipline in the former Soviet empire after genetics), which have taken place all over the Marxist-dominated world. Particularly dangerous was the accusation that cosmology is either a scientific mask for religion, or itself is similar to religion, which could as well cost one one's life in those years. This displays crucial ideological tendency in the anti-cosmological viewpoint. A target for religious dogmatism and fanaticism in the Middle Ages, cosmology in the XX century has lived through a monstrous history of totalitarian repression. In both German Nazism and Soviet Marxism, it was a science beyond the pale, with all appropriate and largely tragic consequences. The most sordid example of that disgusting history is the heinous murder of Matvei Bronshtein in 1938. by Soviet socialists under some nonsensical pretext. Is that just a coincidence, or there is something deeply human and liberating in the cosmological endevoar *per se*, something which necessarily disturbs all totalitarian, closed-minded and self-centered worldviews?

---

[3] Quite apart from the plain fact that Aristotle himself has been rather empirically minded philosopher/scientist. He was much more empirically oriented than, say, Descartes, Newton or Kepler.

## 5. Conclusion: patient alive and well

What else is there in the Disney's criticism apart from issues we have addressed here and some strong-worded rethorics? Rutherford's dictum quoted ("don't let me hear anyone use the word 'Universe' in my department") testifies only on regrettable closed-mindedness, authoritarianism and arrogance to which some directors and managers are sometimes prone. Comparisons with religion abound, but—as we have seen in the previous section—they are not new, and seem to be rather ill motivated ("rapt audience, the media exposure, the big book-sale", as if an element of jealousy is not entirely absent). It is, in absence of precise sociological studies, rather questionable that cosmology receives more attention from non-scientists than do other astronomical disciplines or sciences in general, for instance planetary sciences or searching for extraterrestrial life or genetic engineering. But it is one thing to criticize a phenomenon like the perceived media attention; it is much more sinister to suggest that "<u>for that reason alone</u> [underlined by M. M. Ć.] other scientists simply must treat the pretensions of cosmology... with heightened scepticism." We fail to see the causal connection here. While reiterating the necessity of scepticism **under any circumstances**, the criterion of Disney is simply devoid of rational grounds. *Per analogiam*, one expects that such disciplines as computer science, astronautics or genetics should be treated with heightened scepticism because they are in the focus of public attention in recent years (and, parenthetically, much more so than cosmology). Besides, it is rather uncertain what does this "heightened scepticism" really mean. Does this mean that professional journals in genetics should have systematically more referees than journals in botanic? Or that sales of books in artificial intelligence or cosmology ought to be administratively limited?

Finally, the message to the general public which the discussed article suggests has some elements of a tautology. The points that our theories are changeable and replaceable is nothing new or particularly interesting. However, minor slips of dogmatism are present here also (e.g. "the architecture and history of <u>infinity</u>" [underlined by M. M. Ć.]).

In conclusion, the return of critiques of cosmology on the broadest basis is an interesting phenomenon for the sociologists and philosophers of science, but does not present anything new or constructive in physical sense. Such critiques are partially unfair and highly selective, as we have shown above. The most serious problem with such critiques lies, in the opinion of the present author, in the absence of originality. At the close of the century and the millenium, one may conclude that the attacks did not inflict much damage, since cosmology as the scientific practice is livelier than ever. However, it tells something on the lack of new motivation and ideas on the part of the detractors.

**Acknowledgements.** The crucial conclusions of the present study occured to the author during discussions with Profs. Petar Grujić and Miloš Arsenijević, who are, therefore, to be wholeheartedly acknowledged together with the entire Alternative Academic Network of Belgrade and its noble mission. Help in obtaining some of the references, as well as kind support has been received from Maja Bulatović, Branislav Nikolić and Vesna Milošević-Zdjelar. The discussions with Nick Bostrom, Mašan Bogdanovski and Srdjan Samurović have been stimulating and helpful in development of the ideas sketched here.